\documentclass[aps, pra, superscriptaddress, reprint]{revtex4-2}
\usepackage{amsmath}
\usepackage{amssymb}
\usepackage{braket}
\usepackage{bbm}
\usepackage{graphicx}
\usepackage{tikz}
\usepackage{caption}
\usepackage{subcaption}
\usepackage{hyperref}
\usepackage{verbatim}
\hypersetup{
    colorlinks=true,
    linkcolor=blue,
    filecolor=blue,      
    urlcolor=blue,
    citecolor = blue,
    breaklinks=true,
}

\allowdisplaybreaks

\usepackage{natbib}

\usepackage{array}
\newcolumntype{P}[1]{>{\centering\arraybackslash}p{#1}}

\newcommand{\refI}[1]{{{#1}}}
\newcommand{\refII}[1]{{{#1}}}
\newcommand{\refBoth}[1]{{{#1}}}
\newcommand{\refExt}[1]{{{#1}}}

\newcommand{\spect}{\phi_{\lambda}(\omega)}

\begin{document}
\title{Adiabatic Gauge Potential as a Tool for Detecting Chaos in Classical Systems}
\author{Nachiket Karve}
\email{nachiket@bu.edu}
\author{Nathan Rose}
\author{David Campbell}
\affiliation{Department of Physics, Boston University, Boston, Massachusetts 02215, USA}

\date{\today}

\begin{abstract}
The interplay between chaos and thermalization in weakly non-integrable systems is a rich and complex subject.  Interest in this area is further motivated by a desire to develop a unified picture of chaos for both quantum and classical systems.  In this work, we study the adiabatic gauge potential (AGP), an object typically studied in quantum mechanics that describes deformations of a quantum state under adiabatic variation of the Hamiltonian, in classical Fermi-Pasta-Ulam-Tsingou (FPUT) and Toda models. We show how the time variance of the AGP over a trajectory probes the long-time correlations of a generic observable and can be used to distinguish among nearly integrable, weakly chaotic, and strongly chaotic regimes. We draw connections between the evolution of the AGP and diffusion and derive a fluctuation-dissipation relation that connects its variance to long-time correlations of the observable. Within this framework, we demonstrate that strongly and weakly chaotic regimes correspond to normal and anomalous diffusion, respectively.  The latter gives rise to a marked increase in the variance as the time interval is increased, and this behavior serves as the basis for our probe of the onset times of chaos, \refBoth{which is interpreted as a ``mixing" time.} Numerical results are presented for FPUT and Toda systems that highlight integrable, weakly chaotic, and strongly chaotic regimes. \refBoth{Further, a hierarchy of $t_{\text{Lyapunov}} < t_{\text{chaos}} < t_{\text{thermalization}}$ is found in these models.} We conclude by commenting on the wide applicability of our method to a broader class of systems.
\end{abstract}

\maketitle

\section{Introduction}

Many natural processes exhibit chaotic behavior under deterministic dynamics, even when isolated from external randomness \cite{berry_1978}. Unpredictable motion can arise even in small systems, such as a double pendulum \cite{shinbrot_1992} or the three-body problem \cite{hietarinta_1993}. Classical chaos is characterized by a system’s sensitivity to small variations in its initial conditions. This sensitivity is quantified by Lyapunov exponents, which measure the exponential divergence of initially close trajectories in phase space \cite{oseledets_1968, young_2013}. \refExt{Numerous studies \cite{mccartney_2011, benettin_2018, flach_2022, kevrekidis_2023} have used Lyapunov exponents to characterize chaotic behavior in classical systems, and there exist widely used numerical methods to compute them \cite{benettin_1980, sandri_1996}.} \refI{We note that this discussion is limited to non-relativistic systems since Lyapunov exponents are observer dependent under Lorentz transformations, and thus require more careful treatment in relativistic settings. \cite{hashimoto_2023,szczesny_1999,motter_2009}}

\begin{figure}
    \centering
    \includegraphics[width=0.75\linewidth]{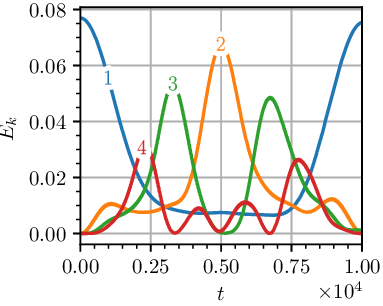}
    \caption{Results of the original FPUT experiment with cubic non-linearities, reproduced from \cite{fermi_1955}. The energies of the first four linear modes of the system are plotted as a function of time (blue, orange, green, and red respectively). The system starts with all of its energy confined to the first mode, and almost all of it returns back periodically.}
    \label{fig_fermi}
\end{figure}

While chaos and ergodicity were thought to arise from non-linearities in classical models, the famous Fermi-Pasta-Ulam-Tsingou (FPUT) experiment apparently challenged that assumption \cite{fermi_1955}. It was observed that a one-dimensional system of non-linearly coupled oscillators, when initialized in a long-wavelength mode, showed recurrent behavior \cite{pierangeli_2018, tuck_1972, pace_2019}. The energy remained confined to only a few modes instead of being distributed equally over the system, and the system periodically returned close to its initial state (See Fig. \ref{fig_fermi} where the original FPUT experiment is reproduced). It is now understood that the FPUT system does indeed thermalize, but over incredibly long times \cite{benettin_2011}. The system goes through a long-lived, non-thermal, ``metastable" phase, akin to a glassy state \cite{kevin_2023,benettin_2008,danieli_2017}. Eventually, the metastable state breaks down, and the FPUT system moves towards equipartition \cite{onorato_2023,lvov_2015}. It is interesting to note that the thermalization time of the original FPUT experiment is thought to be well beyond our current computational capabilities. Studies of the Lyapunov exponents in FPUT systems have suggested the existence of strong and weak chaos regimes \cite{pettini_2005}. 

On the other hand, defining chaos in quantum systems is a harder problem. Quantum mechanics deals with transition probabilities and not trajectories; therefore, it is not clear how to define a counterpart of Lyapunov exponents in quantum systems.  One of the widely recognized definitions of quantum chaos is the Bohigas-Giannoni-Schmit (BGS) conjecture \cite{bgs_1984}, built on Wigner's surmise \cite{wigner_1957} and the Berry-Tabor conjecture \cite{berry_1977}, and states that the energy level statistics of systems with chaotic classical limits are described by random matrices. This was later further generalized to the eigenstate thermalization hypothesis (ETH) \cite{d'alessio_2016,srednicki_1999}. Some other suggested probes of chaos in quantum systems are out-of-time-order correlators (OTOCs) \cite{rozenbaum_2017}, and the growth of operators in Krylov space \cite{parker_2019}.

Recently, Pandey {\it et al.} proposed that quantum chaos can be probed via adiabatic eigenstate deformations \cite{pandey_2020}. They argued that the eigenstates of a chaotic quantum system show complex changes when the system is perturbed adiabatically. The complexity of these deformations is captured by the so-called ``adiabatic gauge potential (AGP)" operator \cite{kolodrubetz_2017,pozsgay_2024}, and the tendency towards chaos is measured via the variance of the AGP over the state space. Further, the scaling of the AGP with the system size has revealed distinct regimes of chaos in many quantum systems \cite{pandey_2020,bhattacharjee_2023,leblond_2021,vidmar_2025,orlov_2023,sharipov_2024}. \refExt{Note that the AGP has previously been studied in the context of counter-diabatic driving and shortcuts to adiabaticity \cite{demirplak_2003,demirplak_2005,demirplak_2008,berry_2009,delcampo_2017,delcampo_2024}.}

Unifying our understanding of chaos in classical and quantum systems is an exciting current challenge. It has been shown that the AGP can be extended and used as a probe in classical spin systems \cite{lim_2024}. Akin to quantum systems, it has been suggested that the behavior of the spectral function's low-frequency tail would reveal the chaotic nature of the classical system. In this work, we take a dynamical approach to the AGP and show that it can indeed be used to study chaos in classical FPUT systems. We claim that this probe allows us to time the transitions of trajectories in the phase space between the near-integrable, weak, and strong chaos regimes. We draw an analogy to diffusion, and derive a fluctuation-dissipation relation describing the growth of the AGP over distributions in the phase space. We generalize this to an ``observable diffusion hypothesis" and show that the different regimes of chaos are related to anomalous diffusion of observables in the phase space of a classical system. The sensitivity of this probe is demonstrated by implementing it in the $\alpha$-FPUT, $\beta$-FPUT, and Toda systems.

The remainder of this paper is organized as follows. We present a brief review of the quantum AGP in Sec. \ref{sec_AGPQuantum}. This is then extended to classical systems in Sec. \ref{sec_AGPclassical}, where we introduce the dynamical fidelity susceptibility. Here, a fluctuation-dissipation relation is derived, and anomalous diffusion of the AGP is related to different regimes of chaos. These ideas are numerically studied in FPUT systems in Sec. \ref{sec_numerics}. Finally, we summarize and discuss our results in Sec. \ref{sec_conclusions}.

\section{Adiabatic Gauge Potential in Quantum Systems}
\label{sec_AGPQuantum}

In this section, we review briefly the AGP in quantum systems. See \cite{kolodrubetz_2017,pandey_2020} for more detailed discussions. Consider a system with Hamiltonian $H(\lambda)$ that depends on a non-linear parameter $\lambda$. The AGP is defined to be the generator of eigenstate deformations under adiabatic changes in $\lambda$:
\begin{equation}
\mathcal{A}_\lambda \ket{n(\lambda)} = i\hbar\partial_\lambda \ket{n(\lambda)},
\end{equation}

\noindent
where the $\ket{n(\lambda)}$ are the instantaneous eigenstates of the system: $H(\lambda)\ket{n(\lambda)} = E(\lambda)\ket{n(\lambda)}$. Non-degenerate perturbation theory allows one to calculate the off-diagonal elements of the AGP:
\begin{equation}
\bra{n}\mathcal{A}_\lambda\ket{m} = \frac{i}{\omega_{mn}} \bra{n}\partial_\lambda H\ket{m},
\end{equation}

\noindent
where $\omega_{mn} = \frac{E_m-E_n}{\hbar}$. There is a gauge freedom in choosing the diagonal elements of the AGP, which are conventionally set to zero. Note that the above expression is ill-defined when there are degeneracies. To eliminate this problem, the AGP is defined with respect to a regularizer:
\begin{equation}
\bra{n}\mathcal{A}_\lambda(\mu)\ket{m} = \frac{i\omega_{mn}}{\omega_{mn}^2 + \mu^2} \bra{n}\partial_\lambda H\ket{m}.
\end{equation}

Alternatively, the above expression can be reformulated as \cite{manjarres_2023}:
\begin{equation}
\mathcal{A}_\lambda(\mu) = \frac{1}{2}\int_{-\infty}^{\infty} d\tau \ \text{sgn}(\tau) e^{-\mu |\tau|} \partial_\lambda H(\tau),
\label{eq_AGPquantum}
\end{equation}

\noindent
with $\partial_\lambda H(\tau) = e^{iH\tau/\hbar}\partial_\lambda H e^{-iH\tau/\hbar}$. This expression is often used to compute the AGP as a function of $\mu$, and its behavior as $\mu\to 0$ is studied \cite{lim_2024}.

Furthermore, it can be shown that the AGP satisfies the following identity:
\begin{equation}
[G_\lambda, H] = 0, \text{ where, } G_\lambda = \partial_\lambda H + \frac{i}{\hbar}[\mathcal{A}_\lambda, H].
\label{eq_AGPcomm}
\end{equation}

In many quantum systems, when the exact AGP cannot be computed, a variational approach is employed to compute an approximate AGP by minimizing the norm of the operator $G_\lambda$ \cite{sugiura_2021,ferreirovelez_2024,lawrence_2025,gjonbalaj_2022,morawetz_2024,xie_2022,hegade_2021}. In terms of the regularizer, this operator can be expressed as:
\begin{equation}
G_\lambda (\mu) = \frac{\mu}{2}\int_{-\infty}^{\infty} d\tau \ e^{-\mu|\tau|} \partial_\lambda H(\tau).
\label{eq_quantumG}
\end{equation}

An object of interest while probing chaos in quantum systems is the fidelity susceptibility (\refExt{also called the norm of the AGP \cite{delcampo_2012,delcampo_2017}}), which is defined as:
\begin{equation}
\chi_\lambda(\mu) = \text{tr}\left\{{A}_\lambda^2(\mu)\rho\right\} - \text{tr}\left\{{A}_\lambda(\mu)\rho\right\}^2,
\label{eq_fidSuscQuantum}
\end{equation}

\noindent
where the average is taken over a suitable probability distribution, $\rho$. We will refer to this fidelity as the ``regularized" fidelity susceptibility, to distinguish it from the ``dynamical" fidelity susceptibility introduced in the next section. This regularized fidelity, $\chi_\lambda$, is often expressed in terms of the spectral function, $\phi_\lambda(\omega)$, as:
\begin{subequations}
\begin{equation}
\chi_\lambda(\mu) = \int_{-\infty}^\infty d\omega \ \frac{\omega^2}{(\omega^2 + \mu^2)^2}\phi_\lambda(\omega),
\label{eq_specFn}
\end{equation}
\begin{equation}
\phi_\lambda(\omega) = \sum_n  \frac{\rho_n}{4\pi}\int_{-\infty}^{\infty} dt \ e^{i\omega t} \bra{n}\left\{\partial_\lambda H(t), \partial_\lambda H(0)\right\}\ket{n}_c,
\end{equation}
\end{subequations}
where $\{\dots \}$ is the anti-commutator, and $\bra{n}\partial_\lambda H(t) \partial_\lambda H(0)\ket{n}_c = \bra{n}\partial_\lambda H(t) \partial_\lambda H(0)\ket{n} - \bra{n}\partial_\lambda H(t) \ket{n}\bra{n}\partial_\lambda H(0)\ket{n}$ is the connected part of the correlation function. Numerical computations of the AGP in certain spin-chains \cite{pandey_2020,leblond_2021} have shown that the low frequency behavior of the spectral function reveals the chaotic nature of the system
\begin{equation}
\phi_\lambda(\omega \to 0) \sim \begin{cases}  0, & \text{ integrable,} \\  \omega^{-1+1/z}, \ z>1, & \text{ intermediate,} \\ \text{constant} > 0, & \text{ ergodic.}   \end{cases}
\label{eq_chaosRegimesQuantum}
\end{equation}

Equivalently, the dependence of the fidelity on the regularizer can be extracted from Eq. \ref{eq_specFn}:
\begin{equation}
\chi_\lambda(\mu \to 0) \sim \begin{cases}  \text{constant}, & \text{ integrable,} \\  1/\mu^{2-1/z}, \ z>1, & \text{ intermediate,} \\ 1/\mu, & \text{ ergodic.}   \end{cases}
\end{equation}

\begin{figure*}
    \centering
    \begin{subfigure}[t]{0.45\linewidth}
        \centering
        \includegraphics[width = \linewidth]{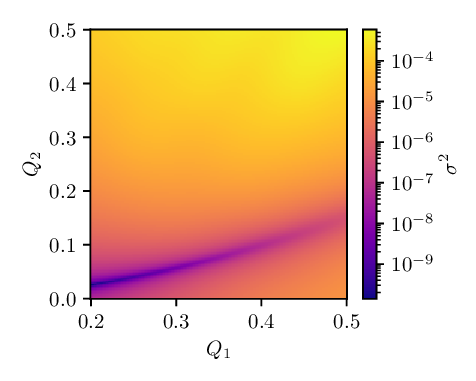}
        \caption{}
    \end{subfigure}
    \hspace{0.1\linewidth}
    \begin{subfigure}[t]{0.36\linewidth}
        \centering
        \includegraphics[width = \linewidth]{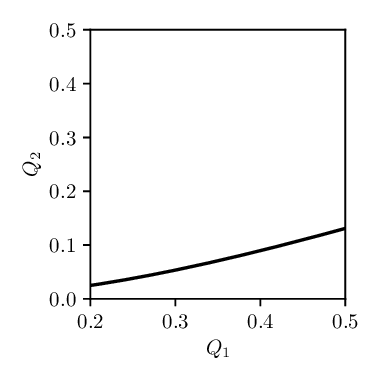}
        \caption{}
    \end{subfigure}
    \caption{(a) The figure depicts a subsection of the phase space of an 8-particle $\alpha$-FPUT system. The variance of the AGP over trajectories starting on this plane, within a time window of $T = 10^{3}$, is plotted as a function of the amplitudes of the first two modes (with all other mode amplitudes and momenta set to zero). There is a sharp drop in the variance along a narrow line in the figure. (b)  Amplitudes of the first two modes of seed mode $k_0 = 1$ $q$-breathers in the $\alpha$-model \cite{flach_2006,flach_2008,karve_2024}. These periodic orbits correspond exactly to the drop in the variance of the AGP. With increasing $T$, the variance is observed to increase everywhere in the phase space, except in the KAM regions.}
    \label{fig_agpPhaseSpace}
\end{figure*}

\section{Classical AGP and Dynamical Fidelity}
\label{sec_AGPclassical}

Much of the discussion of the AGP in quantum systems can be translated to classical systems. The classical AGP is a function of the phase space coordinates and can be defined through a Wigner-Weyl transform of the quantum operator:
\refII{\begin{equation}
\mathcal{A}_\lambda (q,p) = \int dy \ e^{-ipy/\hbar}\bra{q + y/2} \mathcal{A}_\lambda \ket{q - y/2}.
\end{equation}}

The Wigner-Weyl transform maps the quantum AGP operator to a semi-classical function of the phase space coordinates \cite{case_2008}. An expression for the AGP in classical systems is obtained by taking the semiclassical limit of Eq. \ref{eq_AGPcomm}, replacing commutators with Poisson brackets \cite{jarzynski_1995}
\begin{equation}
\{G_\lambda, H\} = 0, \ G_\lambda = \partial_\lambda H -\{\mathcal{A}_\lambda, H\}.
\label{eq_AGPpoisson}
\end{equation}

In integrable systems, the AGP generates canonical transformations that preserve the adiabatic invariants. As shown in \cite{lim_2024}, the Hamiltonian deformation $H(\lambda) \to H(\lambda + \delta \lambda)$ is equivalent to shifting the Hamiltonian by a conserved function $H(\lambda) \to H(\lambda) + \delta\lambda \ G_\lambda$, along with the canonical transformation $q^{\lambda} \to q^{\lambda + \delta\lambda}$, $p^{\lambda} \to p^{\lambda + \delta\lambda}$, given by:
\begin{equation}
\frac{\partial q^\lambda}{\partial \lambda} = -\frac{\partial\mathcal{A}_\lambda}{\partial p^\lambda}, \ \frac{\partial p^\lambda}{\partial \lambda} = \frac{\partial\mathcal{A}_\lambda}{\partial q^\lambda}.
\end{equation}

Thus, the AGP describes deformations to trajectories in the phase space under adiabatic changes. ``Large" deformations under adiabatic changes and, therefore, large variations of the AGP along trajectories, are expected to indicate chaos. We claim that the variance of the AGP over a trajectory, measured in a finite time window, serves as a dynamical probe of chaos along that trajectory:
\begin{equation}
    \sigma^2(T; q,p) = \frac{1}{T}\int_0^T dt \ \mathcal{A}^2_\lambda(t) - \left(\frac{1}{T}\int_0^T dt \ \mathcal{A}_\lambda(t)\right)^2.
\end{equation}
\refII{This trajectory is initialized at the point $(q,p)$ in the phase space and evolved with the Hamiltonian $H(\lambda)$. The variance, $\sigma^2$, is computed over a finite window of time, $T$.} The evolution of the variance can reveal transitions between near-integrable and chaotic regimes and facilitate the computation of thermalization times. As a simple demonstration of its usefulness, see Fig. \ref{fig_agpPhaseSpace}, where trajectories are initialized in a section of the phase space of an $\alpha$-FPUT system, and the variance of the AGP over a time window of $10^3$ is plotted. A sharp drop in $\sigma^2(T; q,p)$ is observed in the vicinity of periodic orbits. We expect ``near-integrable" regions, where trajectories persist for long times, to be found around stable periodic orbits, while chaos is expected in regions further away. \refII{Although the AGP variance in chaotic regions is ``large" compared to near-integrable regions, there is no absolute scale that can be used to determine whether a trajectory is chaotic or not. Rather, as we will show later, whether a region is chaotic or integrable can be determined by how the AGP variance evolves with time: as $T\to\infty$, the AGP variance stays bounded in integrable regions, while it grows without bound in chaotic ones.}

In many cases, averaging the variance of the AGP over an initial, non-stationary probability distribution $\rho(q,p)$ proves useful, as the AGP is often highly sensitive to the initial conditions. This motivates us to define a ``dynamical" fidelity susceptibility as:
\begin{equation}
\chi_\lambda(T) = \int dq \ dp \ \rho(q,p) \ \sigma^2(T; q,p).
\label{eq_fidSuscClassical}
\end{equation}

To compute this dynamical fidelity, we need to obtain the AGP as a function of time along a single trajectory. Using Eq. \ref{eq_AGPpoisson}, the equation of motion for the AGP along a trajectory can be written as:
\begin{equation}
    \frac{d\mathcal{A}_\lambda}{dt} = \{\mathcal{A}_\lambda,H\} = -\partial_\lambda H + G_\lambda.
    \label{eq_AGPeom}
\end{equation}

As shown in appendix \ref{app_AGPclassical}, this equation can be integrated to get:
\begin{equation}
\mathcal{A}_\lambda (t) = \mathcal{A}_\lambda(0) - \int_0^t d\tau \ \left(\partial_\lambda H(\tau) - \overline{\partial_\lambda H}\right),
\label{eq_AGPtraj}
\end{equation}

\noindent
where $\partial_\lambda H(\tau)$ is $\partial_\lambda H$ evaluated at time $\tau$ along the trajectory initialized by $(q,p)$, and $\overline{\partial_\lambda H}$ is its time average over the entire trajectory.
A similar equation is derived in \cite{jarzynski_1995} where the AGP is referred to as the generator of parallel transport. The advantage of this expression is that if the AGP at $t = 0$ is known, then it is very easy to compute the AGP at any other point on the trajectory. In fact, even if the initial AGP cannot be computed, the fidelity susceptibility depends only on the variation of the AGP over the trajectory. Thus, our object of interest is $\Delta \mathcal{A}_\lambda = \mathcal{A}_\lambda (t) - \mathcal{A}_\lambda (0)$, given by:
\begin{equation}
\Delta \mathcal{A}_\lambda (t) =  - \int_0^t d\tau \ \left(\partial_\lambda H(\tau) - \overline{\partial_\lambda H}\right).
\end{equation}

Note that the time average of $\partial_\lambda H$ over a bounded trajectory must converge to a finite value, implying that $\int_0^t d\tau \ \partial_\lambda H(\tau) = t \ \overline{\partial_\lambda H} + C(t)$, where $\lim_{t\to \infty} \frac{C(t)}{t} = 0$. Therefore, $\Delta \mathcal{A}_\lambda(t) = -C(t)$, and the AGP records sublinear deviations of the time integral $\int_0^t d\tau \ \partial_\lambda H(\tau)$. Following Eq. \ref{eq_fidSuscClassical}, this means that the fidelity susceptibility must grow slower than $\mathcal{O}(T^2)$.  

\subsection{AGP Diffusion}
Eq. \ref{eq_AGPtraj} allows us to draw parallels with diffusion processes. Consider an ensemble of Brownian particles, and let $x(t)$ and $v(t)$ denote the position and velocity of an individual particle at time $t$, respectively. By definition,
\begin{equation}
    x(t) = x(0) + \int_0^t d\tau \ v(\tau).
\end{equation}

The above expression has the same form as Eq. \ref{eq_AGPtraj}. The diffusion of Brownian particles is measured through the mean squared displacement, $\Delta x^2(t) = \left\langle \left\{ x(t) - x(0) \right\}^2 \right\rangle$, where $\langle\dots\rangle$ denotes the ensemble average. In contrast, the velocity autocorrelation function $\langle v(t) v(0)\rangle$ is a measure of the fluctuations in the system. The fluctuation-dissipation relation (FDR) relates the spread of the Brownian particles to these fluctuations \cite{kubo_1966}:
\begin{equation}
    \Delta x(t)^2 = t\int_0^{t} d\tau \ \left\langle v(\tau) v(0) \right\rangle.
\end{equation}

When velocities at different times are uncorrelated and the Brownian particle performs a random walk, linear growth of $\Delta x(t)^2$ is observed. This situation is called normal diffusion. On the other hand, anomalous diffusion corresponds to non-linear growth of the mean squared displacement and can result from slow decaying velocity autocorrelations.

Similarly, Eq. \ref{eq_AGPtraj} describes the transport of the AGP over individual trajectories, and the fidelity susceptibility serves as a measure of its mean variance. Under the assumption that the auto-correlation function of $\partial_\lambda H$ depends only on the separation in time, so that
\begin{equation}
\left\langle \left(\partial_\lambda H(t) -\overline{\partial_\lambda H}\right)\left( \partial_\lambda H(t')  -\overline{\partial_\lambda H}\right)\right\rangle  = K(t-t'),
\label{eq_dist}
\end{equation}
where the average $\langle\dots\rangle$ is taken over the distribution $\rho(q,p)$ at $t=0$, a FDR for the AGP can be formulated:
\begin{equation}
\chi_\lambda(T) = \frac{T^2}{6} \int_{-1}^{1} dx \ \left(1-|x|\right)^3 K(xT).
\label{eq_fidSuscCorr}
\end{equation}

See appendix \ref{app_fidSuscClassical} for a detailed derivation of this result. Given $K(t)$, the FDR enables the computation of the dynamical fidelity susceptibility. The long-time behavior of the auto-correlation function, and consequently that of the fidelity susceptibility, is expected to reveal the chaotic nature of the system. Therefore, three distinct regimes can be identified. When the trajectory is highly chaotic, the auto-correlation function $K(t)$ must decay rapidly. Thus, we expect to observe normal diffusion of the AGP \cite{klafter_2005}, and therefore, the strong chaos regime can be defined as corresponding to linear growth of the fidelity. On the other hand, integrable systems remain confined to tori, as described by their integrals of motion. Consequently, we expect the fidelity of near-integrable trajectories to asymptotically approach a finite value. The intermediate regime corresponds to anomalous diffusion, where the growth of the fidelity is non-linear \cite{vlahos_2008}. We will refer to this regime as ``weak chaos". Thus when a state transitions from a near-integrable regime to a chaotic regime (either strong or weak), $\chi_\lambda(T)$ diverges. This facilitates the computation of the ``onset time of chaos". Note the similarity of this proposition to the regularized version \cite{lim_2024}. We compute specific examples in Secs. \ref{subsec_integrable} and \ref{subsec_chaos}, with the results summarized in the table below.
\begin{align}
\bgroup
\def\arraystretch{2.5}
\begin{tabular}{| P{\dimexpr 0.3\linewidth-4\tabcolsep} | P{\dimexpr 0.3\linewidth-3\tabcolsep} | P{\dimexpr 0.3\linewidth-3\tabcolsep} |}
\hline
 &  $K(t\to\infty)$ & $\chi_\lambda(T\to\infty)$ \\ \hline
 Integrable & $\sum\limits_n h_n \cos\omega_n t$ & $\mathcal{O}(T^0)$ \\ \hline
 Weak Chaos & $\sim\dfrac{1}{|t|^{\gamma}}$  $(\gamma \leq 1)$ & $>\mathcal{O}(T)$ \\ \hline
 Strong Chaos & $\sim\delta(t)$, $e^{-|t/t_S|}$, $\dfrac{1}{|t|^{\gamma}}$ $(\gamma > 1)$ & $\mathcal{O}(T)$ \\ \hline
\end{tabular}
\egroup
\end{align}

It is also useful to note the dependence of the fidelity on the spectral function:
\begin{equation}
\chi_\lambda(T) = \int_{-\infty}^{\infty} d\omega \ \frac{\phi_\lambda(\omega)}{\omega^2}\left(1 - \text{sinc}^2\frac{\omega T}{2}\right),
\label{eq_specFnClassical}
\end{equation}

\noindent
where $\text{sinc}\theta$ is the usual sinc function $\frac{\sin\theta}{\theta}$, and $K(t) = \int_{-\infty}^\infty d\omega \ e^{-i\omega t} \phi_\lambda(\omega)$. Compare this with Eq. \ref{eq_specFn}. The integration time $T$ plays the role of the regularizer since it suppresses contributions from the low-frequency tail $\omega << 1/T$.

\refBoth{In general, we expect the time-scale of chaos computed via the norm of the AGP to be larger than the Lyapunov time-scale. While the Lyapunov time characterizes the exponential divergence of infinitesimally close trajectories, finite perturbations can remain bounded for significantly longer periods \cite{aurell_1997}. Therefore, the Lyapunov exponent is not a reliable indicator of the timescales over which trajectories truly ``mix" in the phase space. In contrast, mixing timescales can be more accurately captured by the onset of growth in the fidelity susceptibility, which signals that trajectories are transitioning from an integrable region to other, more chaotic regions of the phase space.}

In fact, all the arguments made in this section also apply to any general observable of the system, and chaos can be probed via ``observable diffusion". To study chaos in a system with Hamiltonian $H$, consider adding a perturbation $\lambda O$, where $O$ is some observable; so that $H(\lambda) = H + \lambda O$ \cite{leblond_2021}. The AGP along a trajectory with respect to this perturbation is simply:
\begin{equation}
    \mathcal{A}_{O}(t) = \mathcal{A}_{O}(0) - \int_0^t d\tau \ \left(O(t) -\overline{O}\right),
\end{equation}
and the FDR describes the diffusion of the AGP associated with the observable $O$:
\begin{equation}
    \chi_O(T) = \frac{T^2}{6} \int_{-1}^{1} dx \ \left(1-|x|\right)^3 \langle O(xT) O(0)\rangle_c.
\end{equation}

\refII{Assuming the observable $O$ to be a smooth function of the phase space variables $(q,p)$, we can state the following. In integrable systems, the autocorrelation function $\langle O(t)O(0)\rangle_c$ must be quasi-periodic, since all trajectories are quasi-periodic. Thus, the AGP variance of any observable in an integrable system must approach a finite value in the long-time limit. Conserved quantities are special, since their AGP variance is always zero. This is true in chaotic systems as well; an observable such as the total energy itself is not a good candidate for probing chaos. Almost all other observables in chaotic systems are expected to have a growing AGP variance, but whether the associated time-scale of chaos is the same for all observables in a given system cannot be said.}

\subsection{Integrable Systems}
\label{subsec_integrable}

All motion in an integrable system is quasi-periodic, and therefore, the auto-correlation function can be written as a Fourier series:
\begin{equation}
    K(t) = \sum_n h_n \cos\omega_n t.
    \label{eq_corrFnIntegrable}
\end{equation}

Here, $\omega_n$s are some linear combinations of the action-angle frequencies. We assume that the frequencies are incommensurate, and that $K(t)$ has no zero-frequency component. We compute the fidelity from Eq. \ref{eq_fidSuscCorr}, and find that:
\begin{equation}
    \chi_\lambda(T) = \sum_n \frac{h_n}{\omega_n^2} + \frac{2h_n}{\omega_n^4 T^2} \left[ \cos\omega_n T - 1 \right].
    \label{eq_fidSuscIntegrable}
\end{equation}

As expected, when $T \to \infty$, the second term in the above equation vanishes, and the fidelity approaches a constant value.

\subsection{Strong and Weak Chaos}
\label{subsec_chaos}

Correlations in non-integrable systems are expected to decay over time. However, certain systems (such as the FPUT system) are known to go through long intermediate phases, where the motion appears to be quasi-periodic. We consider a simple model of such a system which appears integrable below a cut-off time $t_c$, but the correlation decays with a power law at longer times:
\begin{equation}
K(t) = \begin{cases}\sum\limits_n h_n \cos\omega_n t, \ & t < t_c, \\
\frac{h}{|t|^\gamma}, \ & t > t_c. \end{cases}
\end{equation}

The fidelity susceptibility at times $T >> t_c$ can be computed using Eq. \ref{eq_fidSuscCorr}, and the leading order term is given by:
\begin{equation}
\chi_\lambda(T\to\infty) \to \begin{cases} C_1 T^{2-\gamma}, \ & \gamma < 1, \\
C_2 T\ln T, \ & \gamma = 1, \\ 
 C_3 T, \ & \gamma > 1,  \end{cases}
\end{equation}
where $C_1 = \frac{2h}{(1-\gamma)(2-\gamma)(3-\gamma)(4-\gamma)}$, $C_2 = \frac{h}{3}$, and $C_3 = \left[\sum\limits_n \frac{h_n}{3\omega_n}\left(\sin\omega_n t_c-1\right)+\frac{h}{3t_c^{\gamma-1}(\gamma-1)}\right]$. Thus, the system exhibits weak chaos for $\gamma \leq 1$, and strong chaos otherwise.

A strongly chaotic system can also have an exponentially decaying correlation function: $K(t\to\infty) \to h e^{-|t/t_S|}$, where $t_S$ is a measure of the correlation time. The fidelity grows linearly with time in this case as well:
\begin{equation}
\chi_\lambda(T\to\infty) \to \frac{h t_S}{3} T.
\label{eq_strong}
\end{equation}

The slope of the fidelity contains information about the strength of chaos. Stronger chaos leads to a shorter correlation length, resulting in a smaller slope.

\section{Numerical Computations in FPUT Systems}
\label{sec_numerics}

Before presenting our numerical results in FPUT systems, we provide a brief review here of these and related systems. Consider a system of $N$ particles described by the Hamiltonian:
\begin{equation}
    H = \sum_{n = 1}^{N} \frac{1}{2}p_n^2 + \sum_{n=0}^N V(q_{n+1}-q_{n}),
\end{equation}
where fixed boundary conditions are assumed ($q_0=q_{N+1}=0$). Here, $V(x)$ describes interactions between neighbors, and $V_{\alpha\text{-FPUT}}(x) = \frac{1}{2}x^2 + \frac{1}{3}\alpha x^3$, $V_{\beta\text{-FPUT}}(x) = \frac{1}{2}x^2 + \frac{1}{4}\beta x^4$, and $V_{\text{Toda}}(x) = \frac{1}{(2\alpha)^2}\left[e^{2\alpha x} - 2\alpha x - 1\right]$ correspond to the $\alpha$-FPUT, $\beta$-FPUT \cite{fermi_1955} and Toda systems \cite{toda_1967}, respectively. Note that the Toda system is integrable \cite{toda_2012} and is parametrized by the non-linearity $\alpha$. In fact, for small $\alpha$, $V_{\text{Toda}}(x) = V_{\alpha\text{-FPUT}}(x) + \mathcal{O}(\alpha^2)$, and the Toda model can approximate the short-time behavior of the $\alpha$-FPUT system. This fact has been used in \cite{kevin_2023} to probe the breakdown of the metastable state in the $\alpha$-model.

The linear normal modes are introduced by the following canonical transformation:
\begin{equation}
    \begin{pmatrix} Q_k \\ P_k \end{pmatrix} = \sqrt{\frac{2}{N+1}}\sum_{n=1}^N \sin\left(\frac{\pi n k}{N + 1}\right)\begin{pmatrix} q_n \\ p_n \end{pmatrix},        
\end{equation}
where $Q_k$ and $P_k$ are the mode amplitude and momentum of the $k$th linear mode. The $\alpha$- and $\beta$-FPUT Hamiltonians can be written in the mode space as:
\begin{subequations}
\begin{equation}
    H_{\alpha\text{-FPUT}} = \sum_{k=1}^N \frac{P_k^2 + \omega_k^2 Q_k^2}{2} + \frac{\alpha}{3}\sum_{i,j,k=1}^N A_{ijk} Q_iQ_jQ_k,
\end{equation}
\begin{equation}
    H_{\beta\text{-FPUT}} = \sum_{k=1}^N \frac{P_k^2 + \omega_k^2 Q_k^2}{2} + \frac{\beta}{4}\sum_{i,j,k,l=1}^N B_{ijkl} Q_iQ_jQ_kQ_l,
\end{equation}    
\end{subequations}

where $\omega_k = 2\sin\left(\frac{\pi k}{2(N+1)}\right)$, and
\begin{subequations}
    \begin{equation}
        A_{ijk} = \frac{\omega_i\omega_j\omega_k}{\sqrt{2(N+1)}}\sum_{\pm}\left[\delta_{i\pm j\pm k,0} - \delta_{i\pm j\pm k,2(N+1)}\right],
    \end{equation}
    \begin{equation}
        B_{ijkl} = \frac{\omega_i\omega_j\omega_k\omega_l}{2(N+1)}\sum_{\pm}\left[\delta_{i\pm j\pm k\pm l,0} - \delta_{i\pm j\pm k\pm l,\pm 2(N+1)}\right].
    \end{equation}
\end{subequations}

At zero non-linearity the normal mode energies $E_k = \frac{P_k^2 + \omega_k^2 Q_k^2}{2}$ are conserved independently. The non-linear terms in both Hamiltonians couple multiple normal modes together and facilitate the exchange of energy. The distribution of the energy between the normal modes of the system can be measured by the spectral entropy $S$, defined as:
\begin{equation}
    S = -\sum_{k=1}^N \varepsilon_k \ln\varepsilon_k, \text{ where}, \varepsilon_k = \frac{E_k}{\sum_{k'=1}^N E_{k'}}.
\end{equation}

If all the energy of the system is confined to only a single mode, then $S=0$. On the other hand, in the thermal state we expect the energy to be distributed equally among all modes on average, and therefore, $S\approx\ln N$.

\refII{A natural choice of the non-linear parameter $\lambda$ is $\alpha$ in the $\alpha$-FPUT and Toda models, and $\beta$ in the $\beta$-FPUT model. Therefore, we define the AGP in these systems with respect to the following perturbations:
\begin{subequations}
    \begin{equation}
        \alpha\text{-FPUT: }\  \partial_\lambda H = \frac{1}{3}\sum_{n=0}^N (q_{n+1}-q_n)^3,
    \end{equation}
    \begin{equation}
        \beta\text{-FPUT: }\  \partial_\lambda H = \frac{1}{4}\sum_{n=0}^N (q_{n+1}-q_n)^4,
    \end{equation}
    \begin{align}
        \text{Toda: }\  \partial_\lambda H = \sum_{n=0}^N &\frac{1 - e^{2\alpha(q_{n+1}-q_n)}}{2\alpha^3}\notag\\& + \frac{(q_{n+1}-q_n)(1 + e^{2\alpha(q_{n+1}-q_n)})}{2\alpha^2}.
    \end{align}
\end{subequations}}

It is convenient to scale the Hamiltonian by the energy, $H \to H/E$, so that the energy of the new Hamiltonian is simply $1$. In doing so, the phase space coordinates and the non-linearities must also be scaled: $Q_k \to Q_k/\sqrt{E}$, $P_k \to P_k/\sqrt{E}$, $\alpha \to \alpha\sqrt{E}$, and $\beta\to \beta E$. Thus, the dynamics of the FPUT system depend only on the non-linear parameters $E\alpha^2$ and $E\beta$. It should be noted that the AGP can be rescaled as $\mathcal{A}_\lambda \to \mathcal{A}_\lambda/E^{3/2}$ in the $\alpha$-FPUT and Toda systems, and as $\mathcal{A}_\lambda \to \mathcal{A}_\lambda/E^{2}$ in the $\beta$-FPUT system. \refII{All of our numerical simulations in the rest of this paper have been performed in systems with $E=1$, which allows us to study this rescaled AGP as a function of $E\alpha^2$ and $E\beta$.}

We employed fourth-order and sixth-order symplectic Runge-Kutta integrators \cite{calvo_1993,okunbor_1994} in our simulations. \refII{Lyapunov times were calculated by evolving tangent vectors using standard QR decomposition and taking the inverse of the maximal Lyapunov exponent \cite{benettin_1980,geist_1990}. On the other hand, thermalization times were measured from the saturation of the spectral entropy, as described in \cite{lvov_2015}. Both time scales have been computed and compared in \cite{lando_2025}.}

\begin{figure}
\centering
\begin{subfigure}[t]{0.45\linewidth}
\centering
\includegraphics[width=\linewidth]{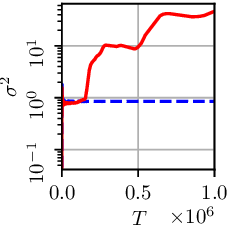}
\caption{}
\end{subfigure}
\hspace{0.05\linewidth}
\begin{subfigure}[t]{0.45\linewidth}
\centering
\includegraphics[width=\linewidth]{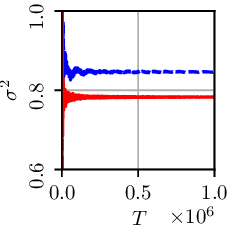}
\caption{}
\end{subfigure}
\caption{The variance of the AGP over individual trajectories in the $\alpha$-FPUT (left) and Toda (right) models with $8$ particles. The plots correspond to non-linearities $E\alpha^2 = 0.090$ (dashed blue line) and $E\alpha^2 = 0.096$ (solid red line). Each trajectory starts in the first mode with $Q_1(0) = \frac{\sqrt{2E}}{\omega_1}$, and $P_1(0) = 0$. Note that in the $\alpha$-model, although the metastable state with the larger non-linearity breaks down at $T \approx 0.2\times 10^6$, the system goes through multiple integrable regions.}
\label{fig_fidSuscIndTraj}
\end{figure}

\subsection{\texorpdfstring{$\alpha$}{alpha}-FPUT and Toda Systems}

\begin{figure*}
\centering
\begin{subfigure}[t]{0.45\linewidth}
    \includegraphics[width = \linewidth]{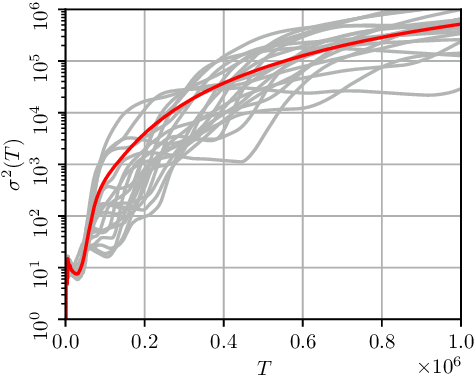}
    \caption{}
    \label{fig_fidSuscAvg}
\end{subfigure}
\hspace{0.05\linewidth}
\begin{subfigure}[t]{0.45\linewidth}
    \includegraphics[width = \linewidth]{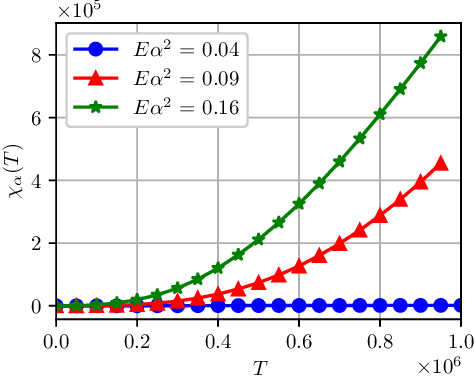}
    \caption{}
    \label{fig_fidSuscWeakAlpha}
\end{subfigure}
\caption{(a) Variance of the AGP over trajectories with different initial phases (gray) in a 32 particle $\alpha$-FPUT system with $E\alpha^2 = 0.09$. The initial condition of each trajectory is $Q_1(0) = \frac{\sqrt{2E}}{\omega_1}\cos\phi$, and $P_1(0) = \sqrt{2E}\sin\phi$. The red line is the average over the phase $\phi$. (b) Phase-averaged fidelity susceptibilities at different non-linearities in a 32-particle $\alpha$-FPUT system. Transitions from the near-integrable to the weak chaotic regime are observed. The growth rate of the fidelity increases with the non-linearity.}
\end{figure*}

\begin{figure*}
\centering
\begin{subfigure}[t]{0.45\linewidth}
\centering
\includegraphics[width = \linewidth]{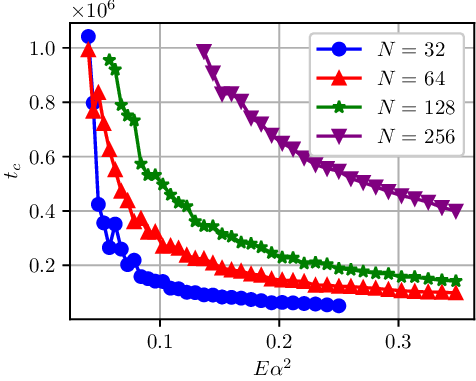}
\caption{}
\label{fig_chaosTime}
\end{subfigure}
\hspace{0.08\linewidth}
\begin{subfigure}[t]{0.45\linewidth}
\centering
\includegraphics[width = \linewidth]{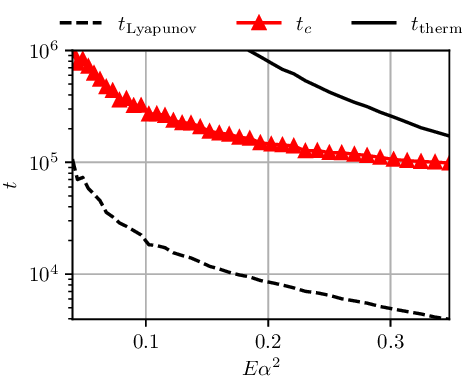}
\caption{}
\label{fig_thermTime}
\end{subfigure}
\caption{(a) Onset time of chaos as a function of non-linearity in the $\alpha$-FPUT system with sizes $N = 32$, $64$, $128$, $256$. (b) \refBoth{Comparison between the Lyapunov time, thermalization time, and the time at which the system enters a chaotic region in a $64$ particle $\alpha$-FPUT system.}}
\end{figure*}

In the original FPUT experiment, the system was initialized in a long-wavelength state, and recurrent motion was observed \cite{fermi_1955}. It is now understood that this almost integrable behavior of the FPUT system is a result of its getting trapped in a long-lived, non-thermal, ``metastable" state \cite{benettin_2008,kevin_2023}. The metastable state slowly moves towards equilibrium, and it has been suggested that the life-time of this state has a power law dependence on the non-linearity \cite{lvov_2015,onorato_2023}. This metastable dynamics is reflected in the AGP variance over such trajectories. As shown in Fig. \ref{fig_fidSuscIndTraj}, $\sigma^2(T;q,p)$ over individual trajectories in the $\alpha$-model, \refII{with initial conditions $Q_k(t=0) = \frac{\sqrt{2}}{\omega_1}\delta_{k,1}$ and $P_k(t=0) = 0$}, appears to saturate at intermediate times, but unlike the Toda model, show a sharp divergence once the metastable state breaks down. Moreover, each trajectory goes through multiple ``metastable regions", as suggested by the plateaus in the graph. These subsequent near-integrable regions are almost impossible to identify from the behavior of the spectral entropy, and therefore, only the first region is conventionally referred to as the metastable state. This behavior is reminiscent of L\'evy flights \cite{chechkin_2008}, in which a particle undergoing a random walk remains spatially localized for extended periods but occasionally takes large, abrupt steps.

\begin{figure*}
    \centering
    \begin{subfigure}[t]{0.45\linewidth}
        \centering
        \includegraphics[width=\linewidth]{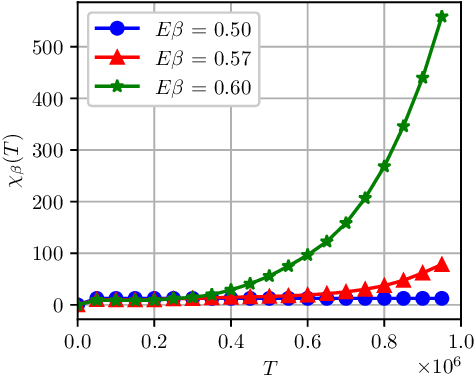}
        \caption{}
        \label{fig_betaRegimesA}
    \end{subfigure}
    \hspace{0.08\linewidth}
    \begin{subfigure}[t]{0.45\linewidth}
        \centering
        \includegraphics[width=\linewidth]{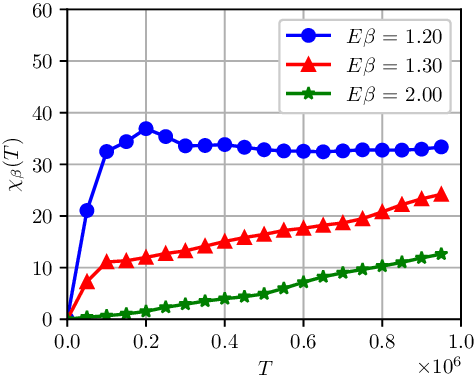}
        \caption{}
        \label{fig_betaRegimesB}
    \end{subfigure}
    \caption{(a) Transition from the near-integrable regime to the weakly chaotic regime in a 32-particle $\beta$-FPUT system. The fidelity susceptibility is constant near integrability, while its growth is faster than linear in the weak regime. The growth rate increases with the non-linearity. (b) Transition from the weak to the strong regime in the same $\beta$-FPUT system. The fidelity grows linearly in the strong regime, and its rate of growth decreases with increasing non-linearity.}
    \label{fig_betaRegimes}
\end{figure*}

\refII{Computing the AGP on trajectories with different initial phases, $Q_k(t=0) = \frac{\sqrt{2}}{\omega_1}\cos\phi \ \delta_{k,1}$ and $P_k(t=0) = \sqrt{2}\sin\phi \ \delta_{k,1}$, we observe a sensitive dependence on the initial condition (Fig. \ref{fig_fidSuscAvg}). Therefore, we find it useful to define the fidelity susceptibility by taking an average of the variance over the phase of the initial mode, $\phi$. The initial distribution is defined so that $\phi$ is uniformly distributed over $[0,2\pi)$. In practice, we average over $1000$ different initial conditions, with $\phi$ chosen randomly from this range.} We see that the phase-averaged fidelity susceptibility varies smoothly with time and shows a power-law growth at large times. \refII{This process is repeated for a wide range of non-linearities, and for each $\alpha$, the initial distribution is evolved independently with $H(\alpha)$.} The rate of growth of the fidelity is observed to increase with the non-linearity (Fig. \ref{fig_fidSuscWeakAlpha}). 

Note that the fidelity appears to saturate to a finite value at intermediate times, indicating that the system is confined to an integrable region. The system is considered to leave this region and enter a chaotic one when $\chi_\alpha(T)$ grows beyond a reasonable threshold. We set this threshold at $\sim 1000$ in our simulations. \refII{The time at which the system enters a chaotic region, $t_c$, is plotted as a function of non-linearity in Fig. \ref{fig_chaosTime}, and a power-law dependence is observed. This onset of chaos is compared to both the Lyapunov time and the thermalization time $t_{\text{therm}}$ in Fig. \ref{fig_thermTime}. As expected, $t_{\text{Lyapunov}} < t_c < t_{\text{therm}}$. The Lyapunov time is a measure of exponential separation of infinitesimally close trajectories; however, $t_c$ being larger than $t_{\text{Lyapunov}}$ indicates that these trajectories are still confined to an integrable region at the Lyapunov time. Beyond $t_c$, the trajectories leave the integrable region and explore the rest of the phase space. They eventually become ergodic and thermalize at $t_{\text{therm}}$.}

\subsection{\texorpdfstring{$\beta$}{beta}-FPUT}

The $\beta$-FPUT model offers an advantage over the $\alpha$-model due to its quartic potential, which is bounded from below and prevents the system from entering a negative potential. This feature enables the study of systems with significantly high non-linearities and facilitates the visualization of strong chaotic regions. Like the $\alpha$-model, we examine phase-averaged fidelity susceptibilities along trajectories initialized in the first mode. Our observations reveal near-integrable, weakly chaotic, and strongly chaotic regions, as illustrated in a 32-particle system in Fig. \ref{fig_betaRegimes}. Weak chaos is observed in systems with $\beta < 1$, and the fidelity exhibits a power-law growth in this regime. The growth rate is observed to increase with the non-linearity. On the other hand, systems with $\beta > 1$ exhibit transitions from the weak to the strong regime. The strongly chaotic regime is characterized by linear growth of the fidelity. Above a certain critical value, $\beta_{\text{crit}}$, the FPUT system bypasses weak chaos entirely. 

\begin{figure}
    \centering
    \includegraphics[width=0.9\linewidth]{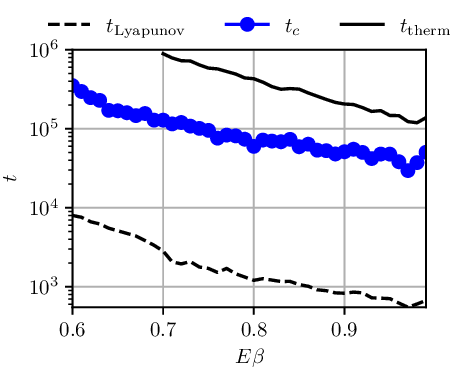}
    \caption{\refBoth{Lyapunov time, thermalization time, and onset time of weak chaos as a function of non-linearity in a 32-particle $\beta$-FPUT model.}}
    \label{fig_thermTimeBeta}
\end{figure}

Unlike the $\alpha$-model, the short-time behavior of the $\beta$-FPUT system cannot be described by a related integrable system. Therefore, it is harder to probe the breakdown of the metastable state in the $\beta$-model using tools developed for the $\alpha$-model \cite{kevin_2023}. However, the ``onset time of weak chaos", $t_c$, can be determined in a manner similar to that used for the $\alpha$-model in the previous section. \refII{See Fig. \ref{fig_thermTimeBeta}, where this time is again compared with the Lyapunov and thermalization times. The system is considered to enter the weakly chaotic regime when the fidelity crosses a threshold of $\sim 100N$. Again, we find that $t_{\text{Lyapunov}} < t_c < t_{\text{therm}}$.}

\begin{figure}
    \centering
    \includegraphics[width=0.9\linewidth]{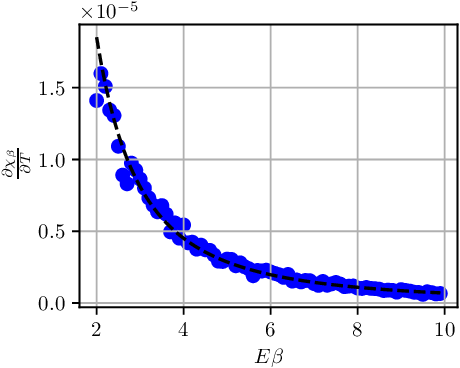}
    \caption{The rate of growth of the fidelity, $\frac{\partial\chi_\beta}{\partial T}$, as a function of the non-linearity in the strong regime of a 32-particle $\beta$-FPUT model. As expected from Eq. \ref{eq_strong}, the growth rate decreases with increasing $\beta$. The slope is observed to decay with a power law: $\frac{\partial\chi_\beta}{\partial T} \propto \left(E\beta\right)^{-2}$.}
    \label{fig_slopes}
\end{figure}

Recall that according to Eq. \ref{eq_strong}, growth of the fidelity in the strong regime is inversely proportional to the strength of chaos. This is evident from Fig \ref{fig_slopes}, which plots the slope of the fidelity in the strong regime. Thus, as $\beta$ is increased above $\beta_{\text{crit}}$, the fidelity crosses the threshold at increasingly later times. Of course, this is not a measure of the ``onset time of strong chaos", since the system starts in the strong regime at $t = 0$, as suggested by the linear growth of $\chi_\beta(T)$ at all times.

\subsection{Spectral Function}

The growth of the fidelity susceptibility is strongly driven by the long-time correlations of observables over the considered time interval, and as such it probes the low-frequency behavior of the spectral function. Previous works studying the fidelity susceptibility in quantum spin chains \cite{pandey_2020}, or both quantum and classical spin systems \cite{lim_2024} have focused heavily on the low-frequency tail of the spectral function.

In our case, we find results consistent with these previous studies, in that the different scaling behaviors of the fidelity susceptibility can be linked to the low-frequency behavior of the spectral function. In particular, the strong growth of the fidelity susceptibility in the weak integrability breaking case follows from the increased spectral weight at the lowest frequencies present in our time window.

\begin{figure}
    \centering
    \includegraphics[width=0.9\linewidth]{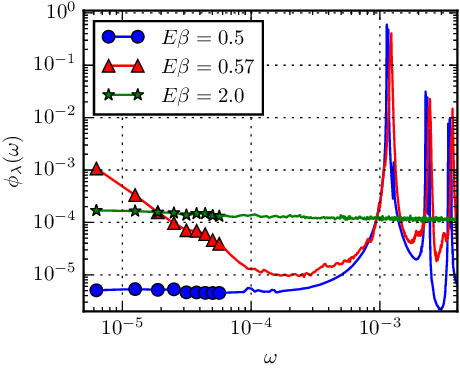} 
    \caption{Low frequency tail of the spectral function for the $\beta$-FPUT system with $N=32$ sites, averaged over $1000$ initial states prepared in the first linear mode with random phases. Increasing values of the non-linearities, $E\beta=0.5,\ 0.57,\ 2.0$ present low frequency behaviors representative of the integrable, weakly chaotic, and strongly chaotic regimes, respectively. Markers omitted at higher frequencies for clarity.}
    \label{fig:beta_spectral_low_tail}
\end{figure}

In Fig. \ref{fig:beta_spectral_low_tail}, the spectral function is plotted for three of the non-linearities shown in Figs. \ref{fig_betaRegimesA}, \ref{fig_betaRegimesB} representing the integrable, weakly chaotic, and strongly chaotic regimes.  For the weakest non-linearity $E\beta=0.5$ the spectral function contains sharp peaks and is small otherwise. Note that in the integrable case, asymptotically the spectral function must vanish no slower than $\spect \sim \omega^2$ as $\omega \rightarrow 0$ in order for the fidelity susceptibility to not grow indefinitely with time. In our case, the spectral function in the integrable regime saturates to a finite but relatively small value. At intermediate nonlinearity $E\beta=0.57$, the spectral function roughly follows that of $E\beta=0.5$ at high frequencies, but deviates prominently as $\omega\rightarrow0$. For $E\beta=2.0$, the spectral function is nearly uniform, indicative of strong chaos. At low frequencies, it approaches a value between the regular and weakly chaotic regimes.

\section{Conclusions}
\label{sec_conclusions}

In this work, we have demonstrated how the AGP can be used as a dynamical probe of chaos in classical systems. Unlike previous studies, where the low frequency behavior of the spectral function and the scaling of the AGP with the system size were used to probe the chaos in the entire system, we have shown that the variance of the AGP on individual trajectories in the phase space can be used to identify transitions between different regimes. By studying the growth of the dynamical fidelity susceptibility, which is the mean variance of the AGP over trajectories in a certain distribution, one can differentiate between near-integrable, strong, and weak chaos regimes. The AGP is observed to ``diffuse" over trajectories, while the fidelity either asymptotically converges to a finite value, grows linearly, or exhibits non-linear growth in the three regimes, respectively. We claim that this probe of chaos can be generalized to an observable diffusion hypothesis, \refII{in other words, we expect the results to hold for generic observables which could serve as perturbations to the Hamiltonian. In this work, we considered perturbations corresponding to varying the nonlinear coupling, but many other choices could be made, and investigating the behavior of the AGP derived from other types of observables would be an interesting topic for future study. }

The method presented in this work provides a general approach to probing chaos. Equation \ref{eq_AGPtraj} enables straightforward computation of the AGP’s evolution along a given trajectory based on the evolution of the perturbation. \refBoth{Compared with Lyapunov exponents, this quantity may be more amenable to measure experimentally, as it can be performed without needing to prepare a nearly identical copy of the system. It is also numerically more efficient, since it does not require the evolution of tangent vectors. Further, the onset of trajectories mixing in the phase space can be identified by a non-linear growth of the fidelity.}

We perform simulations in the $\alpha$-FPUT, $\beta$-FPUT, and Toda systems and probe the breakdown of the metastable state. The transition from near-integrable to weak chaos is timed in the FPUT systems, and a power law dependence on the non-linearity is observed. This transition time is found to be smaller than the thermalization time, but the difference shrinks with increasing non-linearity. As expected, the fidelity grows non-linearly in the weak regime and linearly in the strong one. On the other hand, the fidelity in the Toda system always asymptotically approaches a finite value.

Exploring both discrete and continuous classical systems that exhibit metastable behavior, such as the KdV \cite{zabusky_1965}, mKdV \cite{sal_2019}, and related equations, would be of interest. Additionally, this method can be applied to investigate the emergence of chaos in open systems. Further, this method could be applied to discrete-time maps. Even though such systems cannot be described by a Hamiltonian, the diffusion of a suitable observable can serve as a probe of chaos. Since these maps may exhibit features absent in Hamiltonian systems, such as attractors, repellors, and limit cycles, we anticipate discovering new fidelity behaviors. These questions remain to be addressed in future work.

\section{Acknowledgments}

The authors thank Anatoli Polkovnikov, Bernardo Barrera, Guilherme Delfino, and Adolfo del Campo for insightful discussions. The authors acknowledge the use of Boston University's Shared Computing Cluster (SCC) for numerical simulations.

\section{Data Availability}

The data that support the findings of this article are openly available \cite{karve_git}.

\appendix
\section{AGP Along a Classical Trajectory}
\label{app_AGPclassical}

The classical limits of Eqs. \ref{eq_AGPquantum} and \ref{eq_quantumG} are \cite{lim_2024,berry_1993}:
\begin{subequations}
    \begin{equation}
    \mathcal{A}_\lambda(q,p) = \lim_{\mu \to 0^+}\frac{1}{2}\int_{-\infty}^{\infty} d\tau \ \text{sgn}(\tau) e^{-\mu |\tau|} \partial_\lambda H(\tau),
    \label{eq_AGPclassical}
    \end{equation}
    \begin{equation}
    G_\lambda(q,p) = \lim_{\mu \to 0^+}\frac{\mu}{2}\int_{-\infty}^{\infty} d\tau \  e^{-\mu |\tau|} \partial_\lambda H(\tau),
    \label{eq_Gclassical}
    \end{equation}
\end{subequations}

where $\partial_\lambda H(\tau)$ is $\partial_\lambda H$ evaluated at time $\tau$ along the trajectory initialized by $(q,p)$. Although Eq. \ref{eq_AGPclassical} might not converge as $\mu\to 0$, Eq. \ref{eq_Gclassical} is always finite on bounded trajectories. To see this, we express $G_\lambda(q,p)$ as:
\begin{equation}
    G_\lambda(q,p) = \lim_{\mu \to 0^+}\mu\int_{0}^{\infty} d\tau \  e^{-\mu \tau} \left(\frac{\partial_\lambda H(\tau) + \partial_\lambda H(-\tau)}{2}\right).
\end{equation}

The integral in the above expression $\tilde{H}(\mu) = \int_0^\infty d\tau \ e^{-\mu\tau} \left(\frac{\partial_\lambda H(\tau) + \partial_\lambda H(-\tau)}{2}\right)$ is simply a Laplace transform of $\frac{\partial_\lambda H(\tau) + \partial_\lambda H(-\tau)}{2}$. If we assume that the orbit is bounded (i.e., it does not fall into a negative potential) such that $|\partial_\lambda H(\tau)| < M$, then $\tilde{H}(\mu) < \frac{M}{\mu}$. This means that $\tilde{H}(\mu)$ has no poles on the positive real axis and that $\lim_{\mu\to 0^+} \mu^n \tilde{H}(\mu) = 0$ for $n>1$. Using the final value theorem for Laplace transforms \cite{gluskin_2003}, $G_\lambda$ simply becomes a time average of $\partial_\lambda H$ on the trajectory:
\begin{equation}
    G_\lambda(q,p) = \lim_{T \to \infty}\frac{1}{2T}\int_{-T}^{T} d\tau \  \partial_\lambda H(\tau).
\end{equation}

Therefore, the equation of motion of the AGP, Eq. \ref{eq_AGPeom}, becomes:
\begin{equation}
    \frac{d\mathcal{A}_\lambda}{dt} = \{\mathcal{A}_\lambda,H\} = -\partial_\lambda H + \langle\partial_\lambda H\rangle,
\end{equation}

which can be easily integrated to get Eq. \ref{eq_AGPtraj}.

\section{Classical Fidelity Susceptibility}
\label{app_fidSuscClassical}
To compute the classical fidelity susceptibility, we start with its definition, Eq. \ref{eq_fidSuscClassical}, which is reproduced below:
\begin{align}
\chi_\lambda(T) =& \left\langle\frac{1}{T}\int_0^T dt \ \Delta\mathcal{A}^2_\lambda(t)\right\rangle  \notag\\&\qquad - \left\langle\left(\frac{1}{T}\int_0^T dt \ \Delta\mathcal{A}_\lambda(t)\right)^2\right\rangle,
\end{align}

where the average is taken over the initial distribution $\rho(q,p)$. 
Further, we assume Eq. \ref{eq_dist} for the correlation function of $\partial_\lambda H$. Under these assumptions, let us calculate the following quantity:
\begin{align}
&\left\langle \Delta \mathcal{A}_\lambda(t_1) \Delta \mathcal{A}_\lambda(t_2) \right\rangle \notag\\&= \left\langle \left[-\int_0^{t_1}d\tau_1 \ \left(\partial_\lambda H(\tau_1) - \overline{\partial_\lambda H}\right)\right]\right.\notag\\&\qquad\qquad\times\left.\left[-\int_0^{t_2}d\tau_2 \ \left(\partial_\lambda H(\tau_2) - \overline{\partial_\lambda H}\right)\right] \right\rangle \notag\\&
= \int_0^{t_1}d\tau_1 \int_0^{t_2}d\tau_2 \ K(\tau_1 - \tau_2) \notag\\&
= \int_{-\infty}^\infty d\omega \int_0^{t_1}d\tau_1 \int_0^{t_2}d\tau_2 \ \phi_\lambda(\omega) e^{-i\omega(\tau_1-\tau_2)} \notag\\&
= \int_{-\infty}^\infty d\omega \ \frac{\phi_\lambda(\omega)}{\omega^2} \left(e^{-i\omega t_1} - 1\right) \left(e^{i\omega t_2} - 1\right).
\end{align}

Here, $K(t) = \int_{-\infty}^\infty d\omega \ \phi_\lambda(\omega) e^{-i\omega t}$. And therefore,
\begin{align}
&\left\langle\left(\frac{1}{T}\int_0^T dt \ \Delta\mathcal{A}_\lambda(t)\right)^2\right\rangle \notag\\&= \frac{1}{T^2}\int_0^T dt_1 \int_0^T dt_2 \ \left\langle \Delta \mathcal{A}_\lambda(t_1) \Delta \mathcal{A}_\lambda(t_2) \right\rangle \notag\\&
= \int_{-\infty}^\infty d\omega \ \frac{\phi_\lambda(\omega)}{\omega^2} \left[ \frac{4 \sin^2 \frac{\omega T}{2}}{\omega^2 T^2} - \frac{2\sin\omega T}{\omega T} + 1 \right],
\end{align}

and,
\begin{align}
&\left\langle\frac{1}{T}\int_0^T dt \ \Delta\mathcal{A}^2_\lambda(t)\right\rangle \notag\\&= \frac{1}{T}\int_0^T dt \ \left\langle \Delta\mathcal{A}^2_\lambda(t) \right\rangle \notag\\&
= \int_{-\infty}^\infty d\omega \ \frac{\phi_\lambda(\omega)}{\omega^2} \left[2 - \frac{2\sin\omega T}{\omega T}\right].
\end{align}

Combining the above two equations, we get:
\begin{equation}
\chi_\lambda(T) = \int_{-\infty}^\infty d\omega \ \frac{\phi_\lambda(\omega)}{\omega^2} \left[1 - \text{sinc}^2\frac{\omega T}{2}\right].
\end{equation}

We can plug in $\phi_\lambda(\omega) = \frac{1}{2\pi}\int_{-\infty}^{\infty} dt \ K(t) e^{i\omega t}$ back into the above equation and get:
\begin{align}
\chi_\lambda(T) &= \frac{1}{2\pi}\int_{-\infty}^{\infty} dt \int_{-\infty}^\infty d\omega \ \frac{e^{i\omega t} K(t)}{\omega^2} \left[1 - \text{sinc}^2\frac{\omega T}{2}\right] \notag\\&
= \frac{1}{6T^2}\int_{-T}^{T} dt \ (T-|t|)^3 K(t) \notag\\&
= \frac{T^2}{6} \int_{-1}^{1} dx \ (1-|x|)^3 K(xT).
\end{align}

\bibliographystyle{unsrtnat}
\bibliography{references}

\end{document}